\documentstyle[aps]{revtex}

\begin{document}
\title{Massive gauge field theory without Higgs mechanism}
\author{Jun-Chen Su}
\address{Department of Physics , Jilin University, Changchun 130023, People's\\
Republic of China}
\maketitle

\begin{abstract}
It is argued that the massive gauge field theory without the Higgs mechanism
can well be set up on the gauge-invariance principle based on the viewpoint
that a massive gauge field must be viewed as a constrained system and the
Lorentz condition, as a constraint, must be introduced from the beginning
and imposed on the Yang-Mills Lagrangian. The quantum theory for the massive
gauge fieldis may perfectly be established by the quantization performed in
the Hamiltonian or the Lagrangian path-integral formalism by means of the
Lagrange undetermined multiplier method and shows good renormalizability and
unitarity.
\end{abstract}

It is the prevailing viewpoint that the massive gauge field theory can not
be set up without introducing the Higgs mechanism. The first obstacle is the
gauge-non-invariance of the mass term in the massive Yang-Mills Lagrangian
for a massive gauge field. On the contrary, we present an argument to show
that the conventional viewpoint is not true [1]. In fact, a certain massive
gauge field theory can be well established on the basis of gauge-invariance
principle without recourse to the Higgs mechanism. The basic ideas are
stated in the following.

(1) A massive gauge field must be viewed as a constrained system. In the
previous attempt of building up the massive non-Abelian gauge field theory,
the massive Yang-Mills Lagrangian density written below was chosen to be the
starting point [2,3]. 
\begin{equation}
{\cal L}=-\frac 14F^{a\mu \nu }F_{\mu \nu }^a+\frac 12m^2A^{a\mu }A_\mu ^a 
\eqnum{1}
\end{equation}
where $A_\mu ^a$ are the vector potentials for the non-Abelian massive gauge
fields, 
\begin{equation}
F_{\mu \nu }^a=\partial _\mu A_\nu ^a-\partial _\nu A_\mu ^a+gf^{abc}A_\mu
^bA_\nu ^c  \eqnum{2}
\end{equation}
are the field strengths and $m$ is the mass of gauge bosons. The first term
in the Lagrangian is the ordinary Yang-Mills Lagrangian which is
gauge-invariant under a whole Lie group and used to determine the form of
interactions among the gauge fields themselves. The second term in the
Lagrangian is the mass term which is not gauge-invariant and only affects
the kinematic property of the fields. The above Lagrangian itself was ever
considered to give a complete description of the massive gauge field
dynamics. This consideration is not correct because the Lagrangian is not
only not gauge-invariant, but also contains redundant unphysical degrees of
freedom. As one knows, a full vector potential $A^\mu (x)$ can be split into
two Lorentz-covariant parts: the transverse vector potential $A_T^\mu (x)$
and the longitudinal vector potential $A_L^\mu (x)$, $A^\mu (x)=A_T^\mu
(x)+A_L^\mu (x)$. The transverse vector potential $A_T^{a\mu }(x)$ contains
three independent spatial components which is sufficient to represent the
polarization states of a massive vector boson. Whereas, the longitudinal
vector potential $A_L^{a\mu }$ appears to be a redundant unphysical variable
which must be constrained by introducing the Lorentz condition 
\begin{equation}
\varphi ^a\equiv \partial ^\mu A_\mu ^a=0  \eqnum{3}
\end{equation}
whose solution is $A_L^{a\mu }=0$. With this solution, the massive
Yang-Mills Lagrangian may be expressed in terms of the independent dynamical
variables $A_T^{a\mu }(x)$%
\begin{equation}
{\cal L}=-\frac 14F_T^{a\mu \nu }F_{T\mu \nu }^a+\frac 12m^2A_T^{a\mu
}A_{T\mu }^a  \eqnum{4}
\end{equation}
which gives a complete description of the massive gauge field dynamics. If
we want to represent the dynamics in the whole space of the full vector
potential as described by the massive Yang-Mills Lagrangian in Eq.(1), the
massive gauge field must be treated as a constrained system. In this case,
the Lorentz condition in Eq.(3), as a constraint, is necessary to be
introduced from the onset and imposed on the Lagrangian in Eq.(1) so as to
guarantee the redundant degrees of freedom to be eliminated from the
Lagrangian.

(2) The gauge-invariance should generally be required for the action written
in the physical subspace. Usually, the gauge-invariance is required to the
Lagrangian. From the dynamical viewpoint, as we know, the action is of more
essential significance than the Lagrangian. This is why in Mechanics and
Field Theory, to investigate the dynamical and symmetric properties of a
system, one always starts from the action for the system. Similarly, when we
examine the gauge-symmetric property of a field system, in more general, we
should also see whether the action for the system is gauge-invariant or not.
In particular, for a constrained system such as the massive gauge field, we
should see whether or not the action represented by the independent
dynamical variables is gauge-invariant. This point of view is familiar to us
in the mechanics for constrained systems. Certainly, in some special cases,
the Lagrangian itself is gauge-invariant in the physical subspace so that
the gauge-invariance of the action is ensured. This situation happens for
the massless gauge fields and the massive Abelian gauge field. For the
non-Abelian gauge fields, the infinitesimal gauge transformation usually is
given by [3] 
\begin{equation}
\delta A_\mu ^a=D_\mu ^{ab}\theta ^b  \eqnum{5}
\end{equation}
where 
\begin{equation}
D_\mu ^{ab}=\delta ^{ab}\partial _\mu -gf^{abc}A_\mu ^c  \eqnum{6}
\end{equation}
This gauge transformation is different from the Abelian one in that in the
physical subspace defined by the Lorentz condition, i.e., spanned by the
transverse vector potential $A_T^{a\mu }$, the fields still undergo
nontrivial gauge transformations 
\begin{equation}
\delta A_{T\mu }^a=D_{T\mu }^{ab}\theta ^b  \eqnum{7}
\end{equation}
where 
\begin{equation}
D_{T\mu }^{ab}=\delta ^{ab}\partial _\mu -gf^{abc}A_{T\mu }^c  \eqnum{8}
\end{equation}
Therefore, the mass term in the massive Yang-Mills Lagrangian written in
Eq.(8) is not gauge-invariant. But, the action given by this Lagrangian is
invariant with respect to the gauge transformation shown in Eqs.(7) and (8).
In fact, noticing the identity: $f^{abc}A_T^{a\mu }A_{T\mu }^b=0$ and the
transversity condition (an identity): $\partial ^\mu A_{T\mu }^a=0,$ it is
easy to see 
\begin{equation}
\delta S=-m^2\int d^4x\theta ^a\partial ^\mu A_{T\mu }^a=0  \eqnum{9}
\end{equation}
This shows that the dynamics of massive non-Abelian gauge fields is
gauge-invariant. Alternatively, the gauge-invariance may also be seen from
the action given by the Lagrangian in Eq.(1) which is constrained by the
Lorentz condition in Eq.(3). Under the gauge transformation written in
Eqs.(5) and (6), noticing the identity $f^{abc}A^{a\mu }A_\mu ^b=0$ and the
Lorentz condition, it can be found that 
\begin{equation}
\delta S=-m^2\int d^4x\theta ^a\partial ^\mu A_\mu ^a=0  \eqnum{10}
\end{equation}
This suggests that the massive non-Abelian gauge field theory may also be
set up on the basis of gauge-invariance principle.

(3) Only infinitesimal gauge transformations need to be considered in the
physical subspace. In examining the gauge invariance of the action for the
massive non-Abelian gauge fields, we confine ourself to consider the
infinitesimal gauge transformation only. The reason for this arises from the
fact that the Lorentz condition limits the gauge transformation only to take
place in the vicinity of the unity of the gauge group. In other words, the
residual gauge degrees of freedom existing in the physical subspace are
characterized by the infinitesimal gauge transformations. This fact was
pointed out in the pioneering article by Faddeev and Popov for the
quantization of massless non-Abelian gauge fields [4]. Usually, the dynamics
of massless gauge fields is described by the Yang-Mills Lagrangian. It is
well-known that the Yang-Mills Lagrangian itself is not quantizable, namely,
it can not be used to construct a convergent generating functional of
Green's functions even though it is gauge-invariant with respect to the
whole gauge group. This is because the Yang-Mills Lagrangian contains
redundant unphysical degrees of freedom and hence is not complete for
describing the massless gauge field dynamics unless a suitable constraint
such as the Lorentz condition is introduced to eliminate the unphysical
degrees of freedom. In the article by Faddeev and Popov, the Lorentz
condition is introduced through the following identity [4] 
\begin{equation}
\Delta [A]\int D(g)\delta [\partial ^\mu A_\mu ^g]=1  \eqnum{11}
\end{equation}
which is inserted into the functional integral representing the vacuum-to
vacuum transition amplitude. After doing this , the authors said that '' We
must know $\Delta [A]$ is only for the transverse fields and in this case
all contributions to the last integral are given in the neighborhood of the
unity element of the group''. The delta-functional in Eq.(16) implies $%
\partial ^\mu (A^g)_\mu ^a=\partial ^\mu A_\mu ^a=0$ which represents the
gauge-invariance of the Lorentz condition because the Lorentz condition is
required to hold for all the field variables including the ones before and
after gauge transformations. In the physical subspace where only the
transverse fields are allowed to appear, only infinitesimal gauge
transformations around the unity element are permitted and needed to be
considered in the course of Faddeev-Popov's quantization even though the
Yang-Mills Lagrangian used is invariant under the whole gauge group.
Obviously, there are no reasons of considering the gauge-transformation
property of the fields in the region beyond the physical subspace because
the fields do not exist in that region. By this point, it can be understood
why in the ordinary quantum gauge field theories such as the standard model,
the BRST-transformations are all taken to be infinitesimal.

According to the general procedure, the Lorentz condition (3) may be
incorporated into the Lagrangian (1) by the Lagrange undetermined multiplier
method to give a generalized Lagrangian. In the first order formalism, this
Lagrangian is written as [5] 
\begin{equation}
{\cal L}=\frac 14F^{a\mu \nu }F_{\mu \nu }^a-\frac 12F^{a\mu \nu }(\partial
_\mu A_\nu ^a-\partial _\nu A_\mu ^a+gf^{abc}A_\mu ^bA_\nu ^c)+\frac 12%
m^2A^{a\mu }A_\mu ^a+\lambda ^a\partial ^\mu A_\mu ^a  \eqnum{12}
\end{equation}
where $A_\mu ^a$ and $F_{\mu \nu }^a$ are now treated as the mutually
independent variables and $\lambda ^a$ are chosen to represent the Lagrange
multipliers. Using the canonically conjugate variables defined by 
\begin{equation}
\Pi _\mu ^a(x)=\frac{\partial {\cal L}}{\partial \dot A^{a\mu }}=F_{\mu
0}^a+\lambda ^a\delta _{\mu 0}={\cal \{} 
\begin{tabular}{l}
$F_{k0}^a=E_k^a,$ if $\mu =k=1,2,3;$ \\ 
$\lambda ^a=-E_0^a,$ if $\mu =0,$%
\end{tabular}
\eqnum{13}
\end{equation}
the Lagrangian in Eq.(18) may be rewritten in the canonical form 
\begin{equation}
{\cal L}=E^{a\mu }\dot A_\mu ^a+A_0^aC^a-E_0^a\varphi ^a-{\cal H}  \eqnum{14}
\end{equation}
where 
\begin{equation}
C^a=\partial ^\mu E_\mu ^a+gf^{abc}A_k^bE^{ck}+m^2A_0^a  \eqnum{15}
\end{equation}
\begin{equation}
{\cal H}=\frac 12(E_k^a)^2+\frac 14(F_{ij}^a)^2+\frac 12%
m^2[(A_0^a)^2+(A_k^a)^2]  \eqnum{16}
\end{equation}
here $E_\mu ^a=(E_0^a,E_k^a)$ is a Lorentz vector, ${\cal H}$ is the
Hamiltonian density in which $F_{ij}^a$ are defined in Eq.(2). In the above,
the four-dimensional and the spatial indices are respectively denoted by the
Greek and Latin letters. From Eq.(14), it is clearly seen that the second
and third terms are given respectively by incorporating the constraint
condition

\begin{equation}
C^a=0  \eqnum{17}
\end{equation}
where $C^a$ was represented in Eq.(15) and the Lorentz condition in Eq.(3)
into the Lagrangian.

Now, let us first perform the quantization of the massive non-Abelian gauge
fields in the Hamiltonian path-integral formalism [5]. In accordance with
the general procedure of the quantization, we should first write the
generating functional of Green's functions via the independent canonical
variables which are now chosen to be the transverse parts of the vectors $%
A_\mu ^a$ and $E_\mu ^a$ 
\begin{equation}
Z[J]=\frac 1N\int D(A_T^{a\mu },E_T^{a\mu })exp\{i\int d^4x[E_T^{a\mu }\dot A%
_{T\mu }^a-{\cal H}^{*}(A_T^{a\mu },E_T^{a\mu })+J_T^{a\mu }A_{T\mu }^a]\} 
\eqnum{18}
\end{equation}
where ${\cal H}^{*}(A_T^{a\mu },E_T^{a\mu })$ is the Hamiltonian which is
obtained from the Hamiltonian in Eq.(16) by replacing the constrained
variables $A_L^{a\mu }$ and $E_L^{a\mu }$ with the solutions of equations
(3) and (17) 
\begin{equation}
{\cal H}^{*}(A_T^{a\mu },E_T^{a\mu })={\cal H}(A^{a\mu },E^{a\mu })\mid
_{\varphi ^a=0,C^a=0}  \eqnum{19}
\end{equation}
As mentioned before, Eq.(3) leads to $A_L^{a\mu }=0$. Noticing this solution
and the decomposition $E^{a\mu }(x)=E_T^{a\mu }+E_L^{a\mu }(x)$, when
setting 
\begin{equation}
E_L^{a\mu }(x)=\partial _x^\mu Q^a(x)  \eqnum{21}
\end{equation}
where $Q^a(x)$ is a scalar function, one may get from Eq.(17) an equation
obeyed by the scalar function $Q^a(x)$ 
\begin{equation}
K^{ab}(x)Q^b(x)=W^a(x)  \eqnum{22}
\end{equation}
where 
\begin{equation}
K^{ab}(x)=\delta ^{ab}\Box _x-gf^{abc}A_T^{c\mu }(x)\partial _\mu ^x 
\eqnum{23}
\end{equation}
and 
\begin{equation}
W^a(x)=gf^{abc}E_T^{b\mu }(x)A_{T\mu }^c(x)-m^2A_T^{a0}(x)  \eqnum{24}
\end{equation}
With the aid of the Green's function $G^{ab}(x-y)$ (the ghost particle
propagator) which satisfies the following equation 
\begin{equation}
K^{ac}(x)G^{cb}(x-y)=\delta ^{ab}\delta ^4(x-y)  \eqnum{25}
\end{equation}
one may find the solution to the equation (22) as follows 
\begin{equation}
Q^a(x)=\int d^4yG^{ab}(x-y)W^b(y)  \eqnum{26}
\end{equation}
From the expressions given in Eqs.(21) and (26), one can see that the $%
E_L^{a\mu }(x)$ is a complicated functional of the variables $A_T^{a\mu }$
and $E_T^{a\mu }$ so that the Hamiltonian ${\cal H}^{*}(A_T^{a\mu
},E_T^{a\mu })$ is of much more complicated functional structure which is
not convenient for constructing the diagram technique in the perturbation
theory. Therefore, it is better to express the generating functional in
Eq.(18) in terms of the variables $A_\mu ^a$ and $E_\mu ^a$. For this
purpose, it is necessary to insert the following delta-functional into
Eq.(18) [5] 
\begin{equation}
\delta [A_L^{a\mu }]\delta [E_L^{a\mu }-E_L^{a\mu }(A_T^{a\mu },E_T^{a\mu
})]=detM\delta [C^a]\delta [\varphi ^a]  \eqnum{27}
\end{equation}
where $M$ is the matrix whose elements are defined by the Poisson bracket 
\begin{equation}
\begin{tabular}{l}
$M^{ab}(x,y)=\{C^a(x),\varphi ^b(y)\}=\int d^4x\{\frac{\delta C^a}{\delta
A_\mu ^a(x)}\frac{\delta \varphi ^b}{\delta E^{a\mu }(x)}-\frac{\delta C^a}{%
\delta E_\mu ^a(x)}\frac{\delta \varphi ^b}{\delta A^{a\mu }(x)}\}$ \\ 
$=D_\mu ^{ab}(x)\partial _x^\mu \delta ^4(x-y)$%
\end{tabular}
\eqnum{28}
\end{equation}
The relation in Eq.(27) is easily derived from equations (3) and (17) by
applying the property of delta-functional. Upon inserting Eq.(27) into
Eq.(18) and utilizing the Fourier representation of the delta-functional 
\begin{equation}
\delta [C^a]=\int D(\eta ^a/2\pi )e^{i\int d^4x\eta ^aC^a}  \eqnum{29}
\end{equation}
we have 
\begin{equation}
\begin{tabular}{l}
$Z[J]=\frac 1N\int D(A_\mu ^a,E_\mu ^a,\eta ^a)detM\delta [\partial ^\mu
A_\mu ^a]\exp \{i\int d^4x[E^{a\mu }\dot A_\mu ^a$ \\ 
$+\eta ^aC^a-{\cal H}(A^{a\mu },E^{a\mu })+J^{a\mu }A_\mu ^a]\}$%
\end{tabular}
\eqnum{30}
\end{equation}
In the above exponential, there is a $E_0^a$-related term $E_0^a(\partial
_0A_0^a-\partial _0\eta ^a)$ which permits us to perform the integration
over $E_0^a$, giving a delta-functional 
\begin{equation}
\delta [\partial _0A_0^a-\partial _0\eta ^a]=det|\partial _0|^{-1}\delta
[A_0^a-\eta ^a]  \eqnum{31}
\end{equation}
The determinant $det|\partial _0|^{-1}$, as a constant, may be put in the
normalization constant $N$ and the delta-functional $\delta [A_0^a-\eta ^a]$
will disappear when the integration over $\eta ^a$ is carried out. The
integral over $E_k^a$ is of Gaussian-type and hence easily calculated. After
these manipulations, we arrive at 
\begin{equation}
\begin{tabular}{l}
$Z[J]=\frac 1N\int D(A_\mu ^a)detM\delta [\partial ^\mu A_\mu ^a]exp\{i\int
d^4x[-\frac 14F^{a\mu \nu }F_{\mu \nu }^a$ \\ 
$+\frac 12m^2A^{a\mu }A_\mu ^a+J^{a\mu }A_\mu ^a]\}$%
\end{tabular}
\eqnum{32}
\end{equation}
When employing the familiar expression [4] 
\begin{equation}
detM=\int D(\bar C^a,C^a)e^{i\int d^4xd^4y\bar C^a(x)M^{ab}(x,y)C^b(y)} 
\eqnum{33}
\end{equation}
where $\bar C^a(x)$ and $C^a(x)$ are the mutually conjugate ghost field
variables and the following limit for the Fresnel functional 
\begin{equation}
\delta [\partial ^\mu A_\mu ^a]=\lim_{\alpha \to 0}C[\alpha ]e^{-\frac i{%
2\alpha }\int d^4x(\partial ^\mu A_\mu ^a)^2}  \eqnum{34}
\end{equation}
where $C[\alpha ]\sim \prod_x(\frac i{2\pi \alpha })^{1/2}$ and
supplementing the external source terms for the ghost fields, the generating
functional in Eq.(32) is finally given in the form 
\begin{equation}
Z[J,\overline{\xi },\xi ]=\frac 1N\int D(A_\mu ^a,\bar C^a,C^a)exp\{i\int
d^4x[{\cal L}_{eff}+J^{a\mu }A_\mu ^a+\overline{\xi }^aC^a+\bar C^a\xi ^a]\}
\eqnum{35}
\end{equation}
where 
\begin{equation}
{\cal L}_{eff}=-\frac 14F^{a\mu \nu }F_{\mu \nu }^a+\frac 12m^2A^{a\mu
}A_\mu ^a-\frac 1{2\alpha }(\partial ^\mu A_\mu ^a)^2-\partial ^\mu \bar C%
^aD_\mu ^{ab}C^b  \eqnum{36}
\end{equation}
which is the effective Lagrangian for the quantized massive non-Abelian
gauge field in which the third and fourth terms are the so-called
gauge-fixing term and the ghost term respectively. In Eq.(36), the limit $%
\alpha \to 0$ is implied. Certainly, the theory may be given in arbitrary
gauges $(\alpha \ne 0)$. In this case, as will be seen soon later, the ghost
particle will acquire a spurious mass $\mu =\sqrt{\alpha }m$.

To confirm the result of the quantization given above, let us turn to
quantize the massive non-Abelian gauge fields in the Lagrangian
path-integral formalism. For later convenience, the massive Yang-Mills
Lagrangian in Eq.(1) and the Lorentz constraint condition in Eq.(3) are
respectively generalized to the following forms 
\begin{equation}
{\cal L}_\lambda =-\frac 14F^{a\mu \nu }F_{\mu \nu }^a+\frac 12m^2A^{a\mu
}A_\mu ^a-\frac 12\alpha (\lambda ^a)^2  \eqnum{37}
\end{equation}
and 
\begin{equation}
\partial ^\mu A_\mu ^a+\alpha \lambda ^a=0  \eqnum{38}
\end{equation}
where $\lambda ^a(x)$ are the extra functions which will be identified with
the Lagrange multipliers and $\alpha $ is an arbitrary constant playing the
role of gauge parameter. According to the general procedure for constrained
systems, Eq.(38) may be incorporated into Eq.(37) by the Lagrange
undetermined multiplier method, giving a generalized Lagrangian 
\begin{equation}
{\cal L}_\lambda =-\frac 14F^{a\mu \nu }F_{\mu \nu }^a+\frac 12m^2A^{a\mu
}A_\mu ^a+\lambda ^a\partial ^\mu A_\mu ^a+\frac 12\alpha (\lambda ^a)^2 
\eqnum{39}
\end{equation}
This Lagrangian is obviously not gauge-invariant. However, for building up a
correct gauge field theory, it is necessary to require the action given by
the Lagrangian in Eq.(39) to be invariant with respect to the gauge
transformations shown in Eqs.(5) and (6) so as to guarantee the dynamics of
the gauge field to be gauge-invariant. By this requirement, noticing the
identity $f^{abc}A^{a\mu }A_\mu ^b=0$ and applying the constraint condition
in Eq.(38), we have 
\begin{equation}
\delta S_\lambda =-\frac 1\alpha \int d^4x\partial ^\nu A_\nu ^a(x)\partial
^\mu ({\cal D}_\mu ^{ab}(x)\theta ^b(x))=0  \eqnum{40}
\end{equation}
where 
\begin{equation}
{\cal D}_\mu ^{ab}(x)=\delta ^{ab}\frac{\mu ^2}{\Box _x}\partial _\mu
^x+D_\mu ^{ab}(x)  \eqnum{41}
\end{equation}
in which $\mu ^2=\alpha m^2$ and $D_\mu ^{ab}(x)$ was defined in Eq.(6).
From Eq.(38), we see $\frac 1\alpha \partial ^\nu A_\nu ^a=-\lambda ^a\ne 0$%
. Therefore, to ensure the action to be gauge-invariant, the following
constraint condition on the gauge group is necessary to be required 
\begin{equation}
\partial _x^\mu ({\cal D}_\mu ^{ab}(x)\theta ^b(x))=0  \eqnum{42}
\end{equation}
These are the coupled equations satisfied by the parametric functions $%
\theta ^a(x)$ of the gauge group. Since the Jacobian of the equations is not
singular, $detM\ne 0$ where $M$ is the matrix whose elements are 
\begin{equation}
\begin{tabular}{l}
$M^{ab}(x,y)=\frac{\delta (\partial _x^\mu {\cal D}_\mu ^{ac}(x)\theta ^c(x))%
}{\delta \theta ^b(y)}\mid _{\theta =0}$ \\ 
$=\delta ^{ab}(\Box _x+\mu ^2)\delta ^4(x-y)-gf^{abc}\partial _x^\mu (A_\mu
^c(x)\delta ^4(x-y))$%
\end{tabular}
\eqnum{43}
\end{equation}
the above equations are solvable and would give a set of solutions which
express the functions $\theta ^a(x)$ as functionals of the vector potentials 
$A_\mu ^a(x)$. The constraint conditions in Eq.(42) may also be incorporated
into the Lagrangian in Eq.(39) by the Lagrange undetermined multiplier
method. In doing this, it is convenient, as usually done, to introduce ghost
field variables $C^a(x)$ in such a fashion [3-5]: $\theta ^a(x)=\varsigma
C^a(x)$ where $\varsigma $ is an infinitesimal Grassmann's number. In
accordance with this relation, the constraint condition in Eq.(42) can be
rewritten as 
\begin{equation}
\partial ^\mu ({\cal D}_\mu ^{ab}C^b)=0  \eqnum{44}
\end{equation}
which usually is called ghost equation. When this constraint condition is
incorporated into the Lagrangian in Eq.(39) by the Lagrange multiplier
method, we obtain a more generalized Lagrangian as follows 
\begin{equation}
{\cal L}_\lambda =-\frac 14F^{a\mu \nu }F_{\mu \nu }^a+\frac 12m^2A^{a\mu
}A_\mu ^a+\lambda ^a\partial ^\mu A_\mu ^a+\frac 12\alpha (\lambda ^a)^2+%
\bar C^a\partial ^\mu ({\cal D}_\mu ^{ab}C^b)  \eqnum{45}
\end{equation}
where $\bar C^a(x)$, acting as Lagrange multipliers, are the new scalar
variables conjugate to the ghost variables $C^a(x)$.

At present, we are ready to formulate the quantization of the massive gauge
field in the Lagrangian path-integral formalism. As we learn from the
Lagrange multiplier method, the dynamical and constrained variables as well
as the Lagrange multipliers in Eq.(45) can all be treated as free ones,
varying arbitrarily. Therefore, we are allowed to use this kind of
Lagrangian to construct the generating functional of Green's functions 
\begin{equation}
\begin{tabular}{l}
$Z[J^{a\mu },\overline{\xi }^a,\xi ^a]=\frac 1N\int D(A_\mu ^a,\bar C%
^a,C^a,\lambda ^a)\exp \{i\int d^4x[{\cal L}_\lambda (x)$ \\ 
$+J^{a\mu }(x)A_\mu ^a(x)+\overline{\xi }^a(x)C^a(x)+\overline{C}^a(x)\xi
^a(x)]\}$%
\end{tabular}
\eqnum{46}
\end{equation}
where $D(A_\mu ^a,\cdots ,\lambda ^a)$ denotes the functional integration
measure, $J_\mu ^a$, $\overline{\xi }^a$ and $\xi ^a$ are the external
sources coupled to the gauge and ghost fields and $N$ is the normalization
constant. Looking at the expression of the Lagrangian in Eq.(46), it is seen
that the integral over $\lambda ^a(x)$ is of Gaussian-type. Upon completing
the calculation of this integral, we finally arrive at 
\begin{equation}
\begin{tabular}{l}
$Z[J^{a\mu },\overline{\xi }^a,\xi ^a]=\frac 1N\int D(A_\mu ^a,\bar C%
^a,C^a,)\exp \{i\int d^4x[{\cal L}_{eff}(x)$ \\ 
$+J^{a\mu }(x)A_\mu ^a(x)+\overline{\xi }^a(x)C^a(x)+\overline{C}^a(x)\xi
^a(x)]\}$%
\end{tabular}
\eqnum{47}
\end{equation}
where 
\begin{equation}
{\cal L}_{eff}=-\frac 14F^{a\mu \nu }F_{\mu \nu }^a+\frac 12m^2A^{a\mu
}A_\mu ^a-\frac 1{2\alpha }(\partial ^\mu A_\mu ^a)^2-\partial ^\mu \bar C^a%
{\cal D}_\mu ^{ab}C^b  \eqnum{48}
\end{equation}
is the effective Lagrangian given in the general gauges. In the Landau gauge
($\alpha \rightarrow 0$), The Lagrangian in Eq.(48) just gives the result in
Eq.(36). It has been proved [1] that the above quantization carried out by
means of the Lagrange multiplier method is equivalent to the Faddeev-Popov
approach of quantization [4].

There are three points we would like to emphasize: (1) In the quantization
by the Lagrange multiplier method, the gauge-invariance is always to be
required even in the arbitrary gauge. Moreover, it has been found that the
action given by the Lagrangian in Eq.(48) is invariant under a kind of
BRST-transformations [6]. Thus, the quantum non-Abelian gauge field theory
is set up from beginning to end on the firm basis of gauge-invariance. (2)
In the Lagrangian path-integral formalism, as shown before, the quantized
result is derived by utilizing the infinitesimal gauge transformations. This
result is identical to that obtained by the quantization in the Hamiltonian
path-integral formalism. In the latter quantization, we only need to
calculate the classical Poisson brackets without concerning any gauge
transformation. This fact reveals that to get a correct quantum theory in
the Lagrangian path-integral formalism, the infinitesimal gauge
transformations are only necessary to be taken into account and thereby
confirms the fact that in the physical subspace restricted by the Lorentz
condition, only the infinitesimal gauge transformations are possible to
exist. (3) From the generating functional shown in Eqs.(47) and (48), one
may derive the gauge boson propagator as follows 
\begin{equation}
iD_{\mu \nu }^{ab}(k)=-i\delta ^{ab}\{\frac{g_{\mu \nu }-k_\mu k_\nu /k^2}{%
k^2-m^2+i\varepsilon }+\frac{\alpha k_\mu k_\nu /k^2}{k^2-\mu
^2+i\varepsilon }\}  \eqnum{50}
\end{equation}
which is of good renormalizable behavior. In the zero-mass limit, this
propagator with the massive Yang-Mills Lagrangian and the generating
functional together all go over to the results given in the massless gauge
field theory, different from the quantum theory established previously from
the massive Yang-Mills Lagrangian alone without any constraint [7-10]. For
the previous theory, there occurs a severe contradiction in the zero-mass
limit that the massive Yang-Mills Lagrangian in Eq.(1) is converted to the
massless one, but, the propagator is not and of a singular behavior. In
particular, the previous theory was shown to be nonrenormalizable [3, 7-9]
because the unphysical longitudinal fields and residual gauge degrees of
freedom are not excluded from the theory.

Up to the present, we limit ourself to discuss the gauge fields themselves
without concerning fermion fields. For the gauge fields, in order to
guarantee the mass term in the action to be gauge-invariant, the masses of
all gauge bosons are taken to be the same. If fermions are included,
Obviously, the QCD with massive gluons fulfils this requirement because all
the gluons can be considered to have the same mass. Such a QCD, as has been
proved, is not only renormalizable, but also unitary [6]. The
renormalizability and unitarity are warranted by the fact that the
unphysical degrees of freedom in the theory have been removed by the
constraint conditions in Eqs.(38) and (45). The gauge-fixing term and the
ghost term in Eq.(49) just play the role of counteracting the unphysical
degrees of freedom contained in the massive Yang-Mills Lagrangian as verived
by the perturbative calculations [6].

\section{References}

[1] J. C. Su, Nuovo Cimento 117B (2002) 203.

[2] C. N. Yang and R. L. Mills, Phys. Rev. 96 (1954) 191.

[3] C. Itzykson and F-B, Zuber, Quantum Field Theory, McGraw-Hill, New York
(1980).

[4] L. D. Faddeev and V. N. Popov, Phys. Lett. B25 (1967) 29.

[5] L. D. Faddeev, Theor. Math. Phys., 1 (1970) 1.

[6] J. C. Su, hep.th/9805192; 9805193; 9805194.

[7] H. Umezawa and S. Kamefuchi, Nucl. Phys. 23 (1961) 399.

[8] A. Salam, Nucl. Phys. 18 (1960) 681; Phys. Rev. 127 (1962) 331.

[9] D. G. Boulware, Ann. Phys. 56 (1970) 140.

[10] P. Senjanovic, Ann. Phys. (N.Y.) 100 (1976) 227.

[11] C. Grosse-Knetter, Phys. Rev. D48 (1993) 2854.

[12] N. Banerjee, R. Banerjee and G. Subir, Ann. Phys. (N. Y.) 241 (1995)
237.

\end{document}